\documentclass[final,3p,times,twocolumn]{elsarticle}
 
\usepackage{graphicx}
\usepackage{dcolumn}
\usepackage{bm}
\usepackage{atlasphysics} 
\usepackage[usenames,dvipsnames]{color}                          
\usepackage{amssymb}

\journal{Physics Letters B}

\usepackage{hyperref}
\usepackage[mathlines]{lineno}
                                            
\begin{document}

\title{\vspace{-0.5cm} \flushright{\normalsize{CERN-PH-EP-2011-132}}
  \\ \center{Measurement of the cross section for the production of a
    $W$ boson in association with $b$-jets in $pp$ collisions at
    $\sqrt{s} = 7$ \TeV\ with the ATLAS detector}}

\author{ATLAS Collaboration}

\date{\today}

\begin{abstract}
  A measurement is presented of the cross section for the production
  of a $W$ boson with one or two jets, of which at least one must be a
  $b$-jet, in $pp$ collisions at $\sqrt{s}=7$~\TeV. Production via top
  decay is not included in the signal definition.  The measurement is
  based on 35~\ipb of data collected with the ATLAS detector at the
  LHC. The $W$+$b$-jet cross section is defined for jets reconstructed
  with the anti-$k_t$ clustering algorithm with transverse momentum
  above 25 GeV and rapidity within $\pm 2.1$. The $b$-jets are
  identified by reconstructing secondary vertices. The fiducial cross
  section is measured both for the electron and muon decay channel of
  the $W$ boson and is found to be 10.2 $\pm$ 1.9 (stat) $\pm$ 2.6
  (syst)~pb for one lepton flavour. The results are compared with
  next-to-leading order QCD calculations, which predict a cross
  section smaller than, though consistent with, the measured value.
\end{abstract}

\maketitle

\section{Introduction}
A measurement is presented of the cross section for the production of
a $W$ boson with one or two $b$-jets in proton-proton collisions at a
centre-of-mass energy of 7~TeV. Production via top decay is not
included in the signal definition.  This measurement provides an
important test of quantum chromodynamics (QCD) as it is sensitive to
heavy-flavour quarks in the initial state.  Next-to-leading order
(NLO) QCD predictions for $W$+$b$-jets have made substantial progress
in the last years
\cite{Campbell:2002tg,FebresCordero:2006sj,ref:4FNS5FNSref,ref:4FNS5FNSref2,Frederix:2011qg,Oleari:2011ey},
and now a complete NLO calculation has become available
\cite{ref:4FNS5FNSref3}.  A measurement of the cross section is also
important because $W$+$b$-jet production is a large background to
searches for the Higgs boson in $WH$ production with a decay of $H\to
b \overline{b}$~\cite{bib:ATLASHV1,bib:ATLASHV2}, to measurements of
top quark properties in single~\cite{CSCbook} and pair
production~\cite{Aad:2010ey}, and to searches for physics beyond the
Standard Model~\cite{Goh:2006wj}. A measurement of $W$+$b$-jet
production in proton-antiproton collisions at $\sqrt{s} = 1.96$~TeV by
the CDF Collaboration~\cite{Aaltonen:2009qi} indicates that the
measured cross section is considerably larger than the NLO QCD
predictions.

The $W$+$b$-jet production cross section is measured in the exclusive
1 and 2 jet final states. Jets originating from $b$-quarks (referred
to as $b$-jets) are identified by exploiting the long lifetime and the
large mass of $b$ hadrons. The fiducial cross section is defined at
particle-level by the selection criteria given in
Table~\ref{tab1}~\footnote{ATLAS uses a right-handed coordinate system
  with its origin at the nominal interaction point (IP) in the centre
  of the detector and the $z$-axis along the beam pipe. The $x$-axis
  points from the IP to the centre of the LHC ring, and the $y$ axis
  points upward. Cylindrical coordinates $(r,\phi)$ are used in the
  transverse plane, $\phi$ being the azimuthal angle around the beam
  pipe. The pseudorapidity is defined in terms of the polar angle
  $\theta$ as $\eta=-\ln\tan(\theta/2)$.  The distance $\Delta R$ in
  $\eta$-$\phi$ space is defined as $\Delta R\equiv \sqrt{\Delta\eta^2
    + \Delta\phi^2}$.}. 
The measurement is based on data corresponding to an integrated
luminosity of 35~\ipb and is compared with QCD NLO
predictions~\cite{ref:4FNS5FNS}. A closely related measurement has
been performed, using very similar techniques, in the $Z$+$b$-jet
final state~\cite{Zb}.

\begin{table}[ht]
\centering
\caption{Definition of the phase space for the 
  fiducial cross section. The $W$ transverse mass is defined as
$ m_T = \sqrt{2p^{\ell}_T p^{\nu}_T (1 -
  \cos(\phi^{\ell} - \phi^{\nu})) }$. 
}
\begin{tabular}{l|c}
\hline\hline
Requirement & 
Cut\\ 
\hline \hline
Lepton transverse momentum & $p_T^\ell > 20$ \GeV  \\
Lepton pseudorapidity & $|\eta^\ell|<2.5$  \\
Neutrino transverse momentum &   $p_T^\nu >25$ \GeV \\
$W$ transverse mass & $m_T>40$ \GeV \\
\hline
Jet transverse momentum& $p_T^j>25$ \GeV \\
%\hline
Jet rapidity &  $|y^j|<2.1$   \\
Jet multiplicity & $n \le 2$ \\
$b$-jet multiplicity & $n_b = 1$ or $n_b = 2$\\
\hline
Jet-lepton separation & $\Delta R(\ell\mathrm{,jet}) > 0.5$  \\
\hline
\hline
\end{tabular} 
\label{tab1}
\end{table}

\section{The ATLAS Detector}
The ATLAS detector~\cite{AtlasDetector} consists of an inner detector
tracking system (ID) surrounded by a superconducting solenoid
providing a $2$~T magnetic field, electromagnetic and hadronic
calorimeters, and a muon spectrometer (MS).  The ID consists of pixel
and silicon microstrip detectors inside a transition radiation
tracker.  The electromagnetic calorimeter is a lead liquid-argon (LAr)
detector in the barrel ($|\eta| < 1.475$) and the endcap ($1.375 <
|\eta| < 3.2$) regions.  Hadron calorimetry is based on two different
detector technologies. The barrel ($|\eta| < 0.8$) and extended barrel
($0.8 < |\eta| < 1.7$) calorimeters are composed of
scintillator/steel, while the hadronic endcap calorimeters ($1.5 <
|\eta| < 3.2$) are LAr/copper.  The forward calorimeters ($3.1 <
|\eta| < 4.9$) are instrumented with LAr/tungsten and LAr/copper,
providing electromagnetic and hadronic energy measurements,
respectively. The MS consists of three large superconducting toroids
and a system of three stations of trigger chambers and precision
tracking chambers.

\section{Simulated Event Samples}
Monte Carlo (MC) simulated event samples with full detector
simulation~\cite{AtlasSimulation}, based on the \textsc{Geant}4
program~\cite{Geant4} corrected for all known detector effects, are
used to model the $W$+$b$-jet signal and most of the backgrounds, as
well as to unfold the measured $W$+$b$-jet yield to obtain the
fiducial cross section.

The processes of $W$ boson production in association with $b$-jets,
$c$-jets and light-flavour jets are simulated separately using the
\textsc{Alpgen}~\cite{ALPGEN} generator, interfaced to
\textsc{Herwig}~\cite{Herwig} for parton shower and fragmentation, and
\textsc{Jimmy}~\cite{Jimmy} for the underlying event simulation. The
MLM~\cite{MLM} matching scheme as implemented in \textsc{Alpgen} is
used to remove overlaps between the $n$ and $n + 1$ parton samples
from the matrix element (ME) and the parton shower.  In addition,
overlap between heavy-flavour quarks that originate from ME production
and those that originate from the parton shower is removed.

The $Z$+jets background is simulated
with \textsc{Alpgen} interfaced to \textsc{Herwig} using the same
configuration as for $W$+jets. The diboson ($WW$, $WZ$, $ZZ$)
background is simulated with \textsc{Herwig}. 
The $t$-channel and $Wt$-channel single top processes are simulated
with AcerMC~\cite{AcerMC}, while the $s$-channel process is simulated
with MC@NLO~\cite{MC_at_NLO}.
The inclusive $W$+jets
and $Z$+jets cross sections are normalized to NNLO
predictions~\cite{FEWZ}, and the cross sections of the other
backgrounds are normalized to NLO predictions~\cite{MCFM}. The \ttbar\
background is simulated with MC@NLO interfaced to \textsc{Herwig}. The
\ttbar\ normalisation is extracted from the data.

Multiple interactions per bunch crossing are accounted for by
overlaying simulated minimum bias events. To match the observed
instantaneous luminosity profile, the MC events are reweighted to
yield the same distribution of the number of primary vertices as
measured in the data.

\section{Event Selection}
The analysis is based on the 2010 data set using 35~\ipb of integrated
luminosity with an uncertainty of 3.4\%~\cite{bib:Lum,Aad:2011dr}. The
data are collected using a single electron or muon high $p_T$ trigger.
Trigger thresholds are low enough to ensure that leptons with $p_T >
20$~GeV lie in the efficiency plateau. All events are required to have
a primary vertex that is reconstructed from at least three tracks with
$p_T > 150$~MeV.

Final states are selected with exactly one isolated electron or
muon. Electrons are required to satisfy $E_T > 20$ GeV and $|\eta| <
2.47$.  Electrons in the region between the barrel and the endcap
electromagnetic calorimeters ($1.37 < |\eta| < 1.52$) are removed.  In
addition to the tight selection as defined in Ref.~\cite{incW}, a
$p_T$- and $\eta$-dependent requirement on a combination of
calorimeter and track isolation is designed to yield constant
efficiency and to reduce the large background from multi-jet
production. Muon candidates are constructed from matched ID and MS
tracks and are required to satisfy $p_T > 20$ GeV and $|\eta| <
2.4$. Muons within a distance $\Delta R < 0.4$ of a jet are
rejected. In addition the calorimeter transverse energy and the sum of
track transverse momenta within $\Delta R < 0.3$ of the muon must both
be less than 4 GeV.

Jets are reconstructed using the anti-$k_t$~\cite{Cacciari:2008gp}
algorithm with a radius parameter $R = 0.4$. To take into account the
differences in calorimeter response to electrons and hadrons, a $p_T$-
and $\eta$-dependent factor, derived from simulated events, is applied
to each jet to provide an average energy scale
correction~\cite{JESv16} back to particle level. Events with one or
two reconstructed jets are selected with jet $p_T > 25$ GeV and
rapidity $|y| < 2.1$. All jets within $\Delta R < 0.5$ of a selected
electron are removed. Jets produced in additional interactions are
removed by requiring that 75\% of the sum of the transverse momenta of
the tracks associated to each jet is consistent with originating from
the primary vertex.

The reconstruction of the missing transverse momentum
(\met)~\cite{incW} is based on the energy deposits in calorimeter cells
grouped into three-dimensional clusters. Corrections for
electromagnetic to hadronic energy scale, dead material,
out-of-cluster energy as well as muon momentum for the muon channel
are applied.  The $W$ boson transverse mass ($m_{T}$) is calculated
from the measured lepton momentum, the missing transverse momentum and
the opening angle between the two according to the formula $ m_T =
\sqrt{2p^{\ell}_T p^{\nu}_T (1 - \cos(\phi^{\ell} - \phi^{\nu}))
}$. For both lepton channels \met $> 25$ GeV and $m_{T}$ $> 40$ GeV is
required.

The algorithm used to tag $b$-jets, SV0~\cite{SV0}, is based on the
decay length significance between the primary vertex and the displaced
secondary vertex reconstructed in the jet. Jets with a decay length
significance greater than 5.85 are considered to be $b$-jet
candidates, referred to as $b$-tagged jets. This working point of the
SV0 algorithm ensures about 35\% efficiency for $b$-jets with a mistag
rate of about 0.3\%, and 8\% for light- and $c$-jets, respectively.
The $b$-tagging efficiency is measured in a sample enriched in
$b$-jets by requiring that the jet contains a muon, which is expected
to come predominantly from a semileptonic $b$ hadron
decay~\cite{btagcalib}. The muon momentum relative to the jet axis,
referred to as $p_T^{rel}$, is used to discriminate $b$-jets from $c$-
and light-flavour jets. The ratio of the $b$-tagging efficiency
measured in data and in the MC simulation is applied to the simulated
samples in the form of a correction factor. This correction factor
does not show any strong dependence on jet $p_T$ or $\eta$ and is
consistent with unity. The total uncertainty on the correction factor
ranges from 6\% to 13\%. These results are confirmed with independent
$b$-tagging efficiency measurements in \ttbar\ events and
alternatively using partial reconstruction of $b$ hadrons in jets in
$D^*\mu$ final states~\cite{btagcalib}.

The overall fraction of $W$+$b$-jet events with two $b$-tagged jets is
negligible (2\%).  Most of the $W$+$b$-jet events with two true
$b$-jets are reconstructed as events with one $b$-jet candidate. This
is due to the requirement of central and high $p_T$ jets and to the
b-tagging efficiency of about $~35$\%. In addition, events containing
more than one $b$-jet candidate are predicted to be dominated by
\ttbar.  Therefore events are selected with one and only one tagged
jet despite the measurement also being sensitive to the production of
$W$+$b$-jet with two true $b$-jets.

\section{Background Estimation and Cross Section Extraction}
Charm hadrons also have an appreciable
lifetime which can result in reconstructed displaced secondary
vertices. Light-flavour jets can also be misidentified as $b$-jets due
to hadronic interactions and photon conversions in detector material,
long-lived light-flavour hadrons like $\Lambda$ and \kshort\ and
wrongly reconstructed displaced vertices.  The invariant mass of the
secondary vertex, $m_{SV}$, is correlated with the mass of the parent
hadron and thus discriminates between $b$-, $c$- and light-flavour
jets. The number of $W$+$b$-jet events is extracted from data by
fitting the measured $m_{SV}$ distribution with a linear combination
of templates for $b$-, $c$- and light-flavour jets using a binned
maximum likelihood fit, while the expected contributions from
non-$W$+jets background processes are constrained in the fit using the
estimated template shapes and normalisations. The $m_{SV}$ 
is calculated from the tracks associated to the secondary vertex
   assuming they are pions.
The fit procedure is validated with simulated pseudo-experiments with flavour
compositions and background levels similar to the measured ones.

The non-$W$+jets background sources comprise top quark pair, single
top, multi-jet and the other electroweak (EW) production processes,
$Z$+jets and dibosons.

The \ttbar\ background is estimated from data by applying the same
secondary vertex mass fit to a control region enriched in \ttbar\
using the same event selection except requiring four or more jets
instead of one or two. Backgrounds to the \ttbar\ process are
estimated in the same way as for the fit in the signal region.  The
$W$+$b$-jet contamination in the \ttbar\ control region is at the 5\%
level and is extrapolated from the measured yield in the signal region
by using \textsc{Alpgen} and an uncertainty of $\pm$~100\%.  The
measured \ttbar\ yield, $n_{\ttbar,measured}^{\ge 4 jets}$, is then
projected into the signal region using MC simulation: $n_{\ttbar}^{1,2
  {\rm jets}}=n_{\ttbar{\rm ,measured}}^{\ge 4{\rm jets}}\cdot
\frac{n_{\ttbar{\rm ,expected}}^{1,2 {\rm jets}}}{n_{\ttbar{\rm
      ,expected}}^{\ge 4 {\rm jets}}}$.  This data-driven \ttbar\
yield estimate is in good agreement with MC@NLO prediction and has the
advantage that it is almost completely independent of the $b$-tagging
uncertainty.  The \ttbar\ $m_{SV}$ template is modelled using MC
simulation.

\begin{figure}[h!]
  \centering
  \begin{tabular}{c}
    \includegraphics[width=1.0\columnwidth]{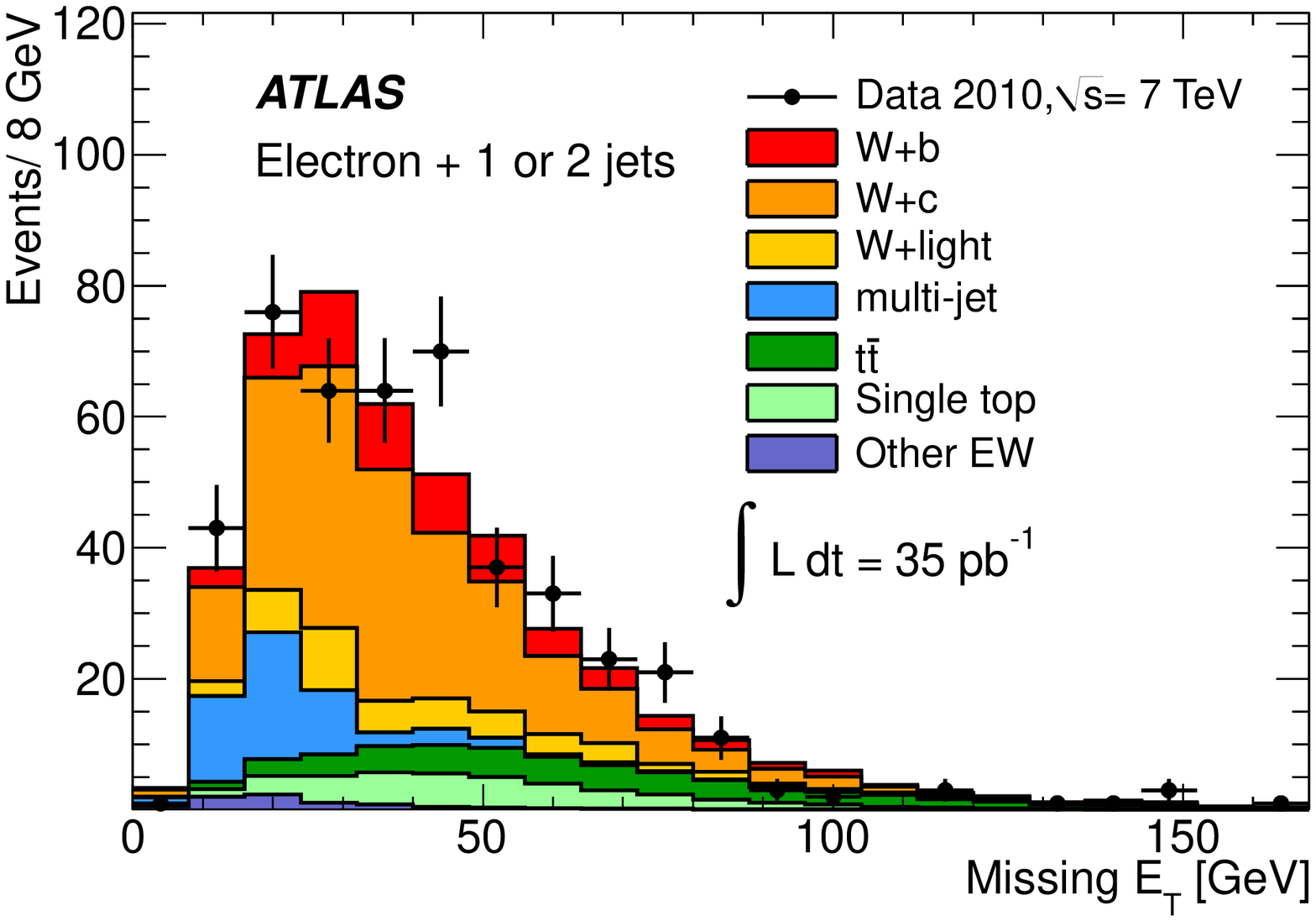} \\
    \includegraphics[width=1.0\columnwidth]{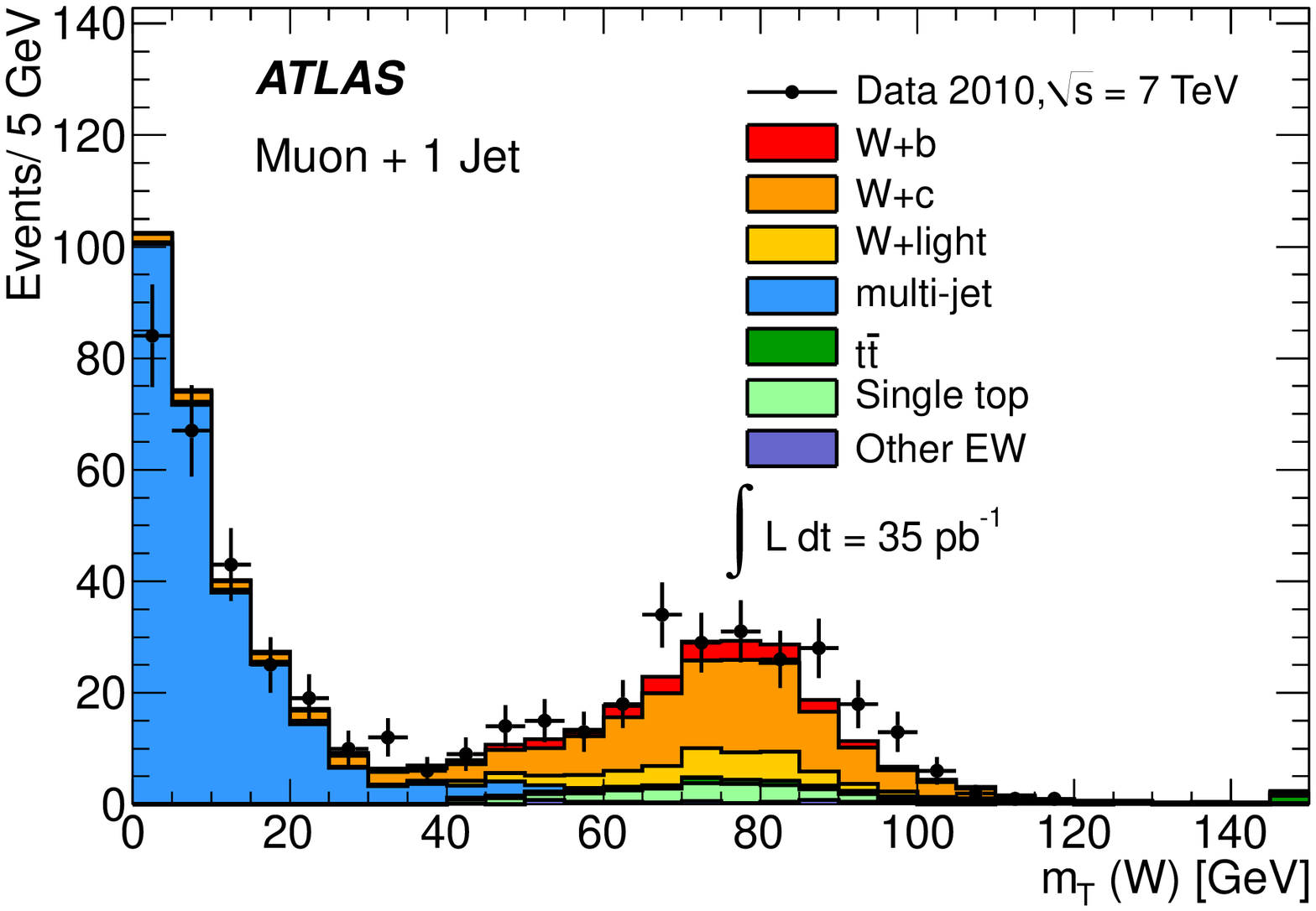} 
  \end{tabular}
  \caption{ (top) \met\ distribution in the electron channel in the
    combined 1- and 2-jet bin without applying the \met\ selection
    criterion. (bottom) $m_{T}$ distribution in the muon channel in
    the 1-jet bin without applying the $m_{T}$ selection criterion
    after applying the $b$-tagging requirement. Non-multi-jet
    contributions are normalized to their MC predictions.  }
  \label{fig1}
\end{figure}

As the multi-jet background is difficult to model with simulation,
data-driven techniques similar to those described in
Ref.~\cite{Aad:2010ey} are used to estimate this background in each
jet multiplicity bin and lepton flavour.  The multi-jet background in
the electron channel arises mainly from non-prompt electrons and a
small amount of fake electrons such as electrons from photon
conversions and misidentified jets with high electromagnetic
fractions.  A binned likelihood template fit of the \met\ distribution
is used to estimate the multi-jet background.  The \met\ template for
multi-jet events is modelled using a complementary data sample where
the full event selection including the $b$-tagging requirement is
satisfied but electrons are required to fail certain selection
criteria and to satisfy a looser identification requirement. This
selection is dominated by multi-jet events. The \met\ template for the
other contributions is modelled using MC simulation. The results of
the fit are shown in Fig.~\ref{fig1}, where the fit region goes from
0-100~GeV. The $m_{SV}$ template is modelled using the same control
region used to model the multi-jet \met\ template.

The muon multi-jet background is dominated by non-prompt muons and
extracted using the matrix method~\cite{Aad:2010ey}. The method is
based on the difference in efficiency for a ‘real’ (prompt) or a
‘fake’ (non-prompt) muon that satisfies a loose selection criterion,
to also satisfy the standard selection criterion.  Fig.~\ref{fig1}
illustrates that the muon multi-jet background is well modelled with
this method. The shape of the $m_{SV}$ template is modelled using a
control region enriched in multi-jet events where the full event
selection including the $b$-tagging requirement is satisfied but
\met$<$10~GeV is required.  For both lepton flavours the multi-jet
background is dominated by real $b$-jets.  The validity of these
approaches to the multi-jet background estimates are verified on
samples of simulated events.

\begin{figure}[h!]
  \centering
  \begin{tabular}{ccc}
    \includegraphics[width=1.0\columnwidth]{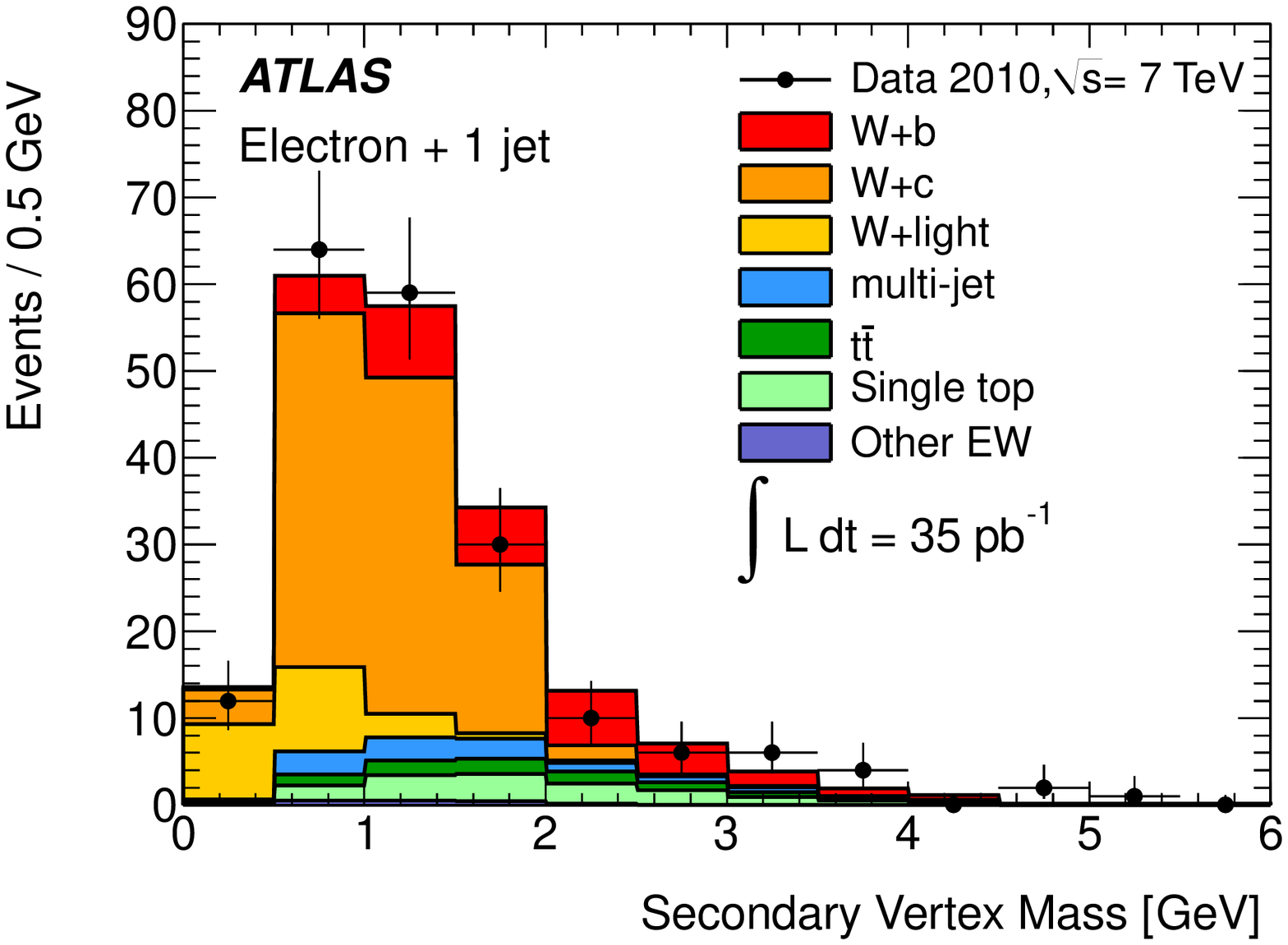} \\
    \includegraphics[width=1.0\columnwidth]{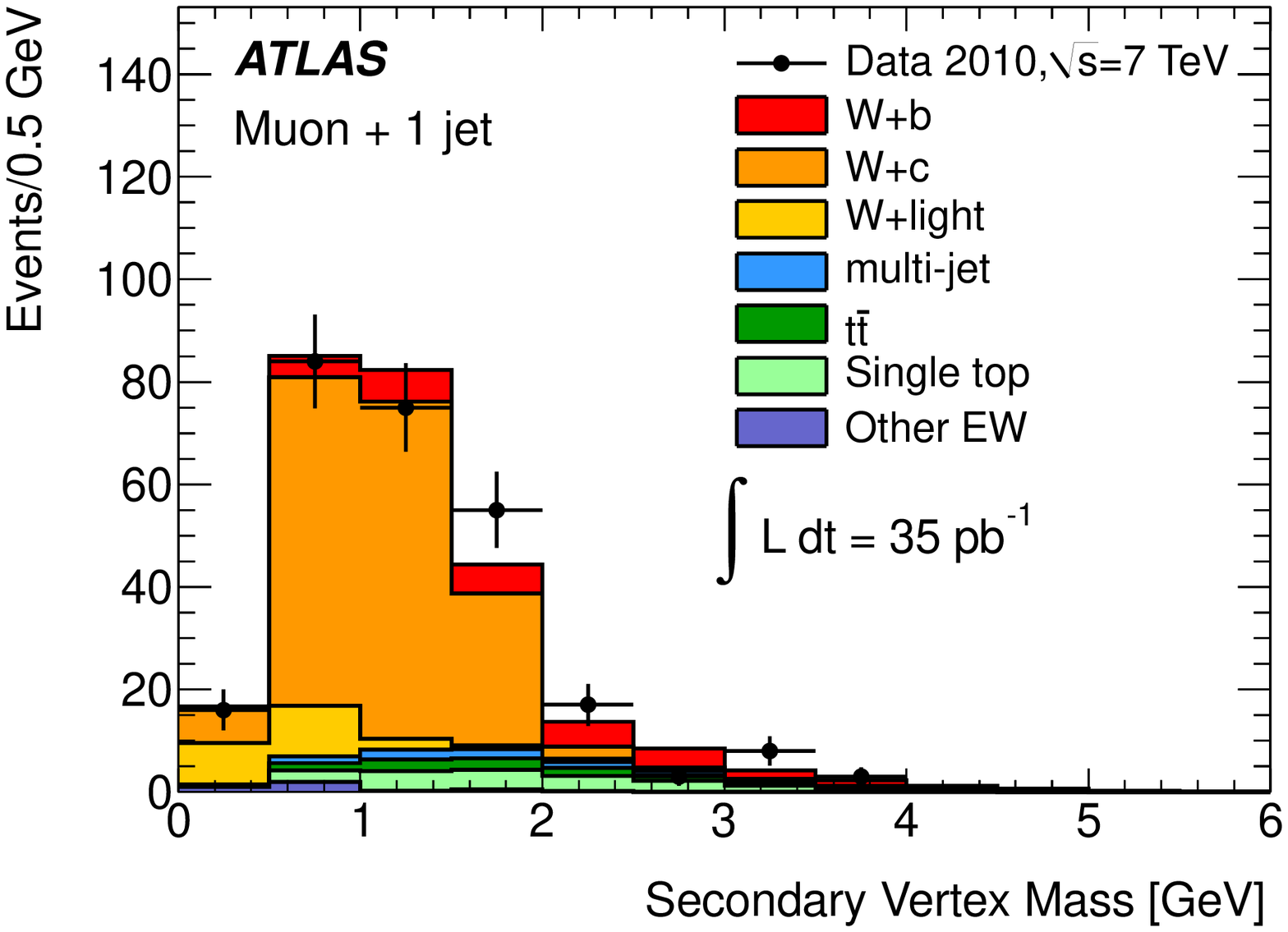} 
  \end{tabular}
  \caption{$m_{SV}$ distributions for the $b$-tagged jet in data and MC,
    where the $W$+jets samples are normalized to the results of the
    maximum likelihood fit and non-$W$+jets backgrounds are normalized
    to the estimates as given in the text, in the 1-jet bin in the
    electron channel (top) and the muon channel (bottom). The stack
    order is the same as in the legend.}
  \label{fig2}
\end{figure}

 \begin{figure}[h!]
   \centering
   \begin{tabular}{ccc}
     \includegraphics[width=1.0\columnwidth]{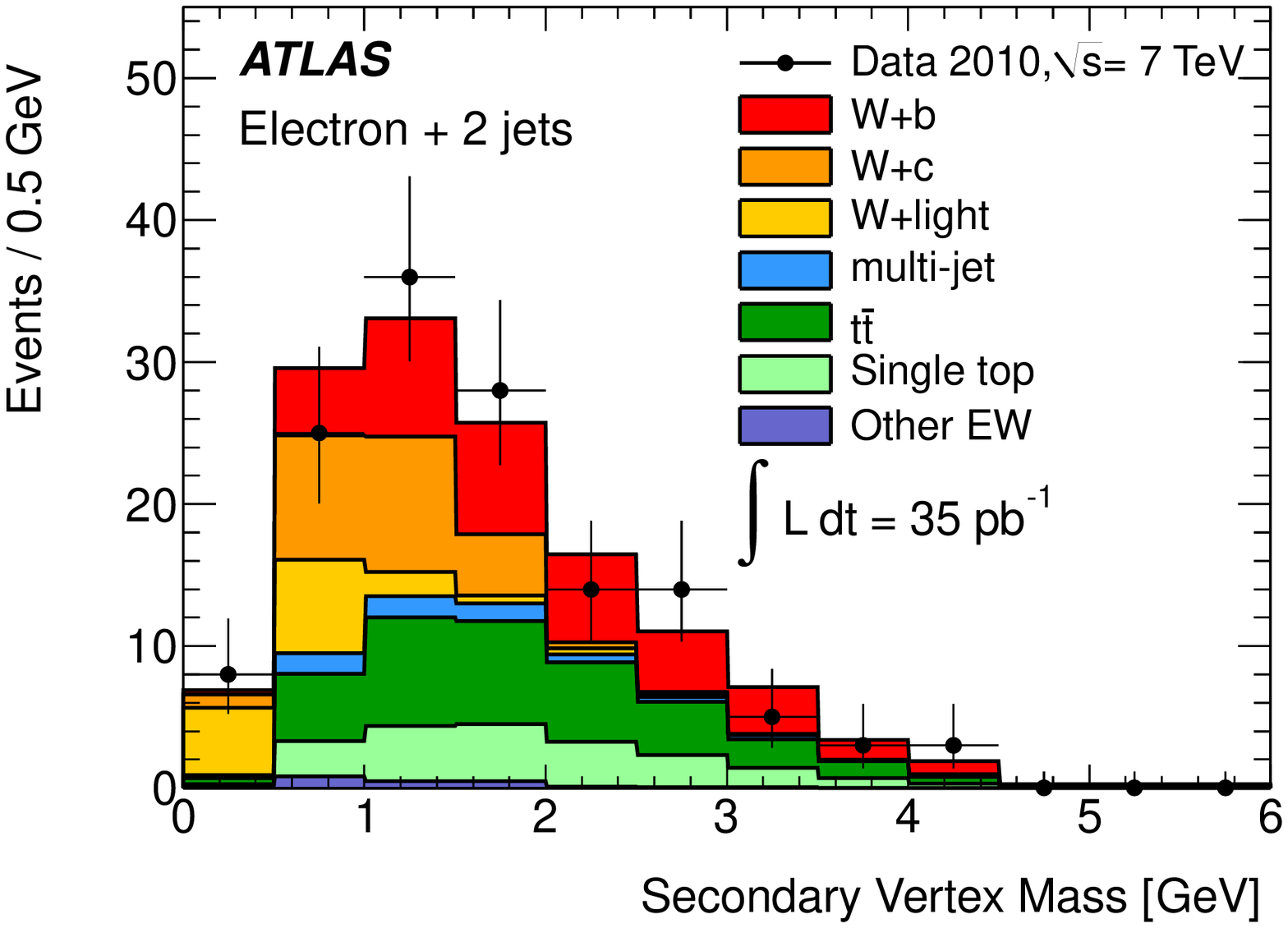} \\
     \includegraphics[width=1.0\columnwidth]{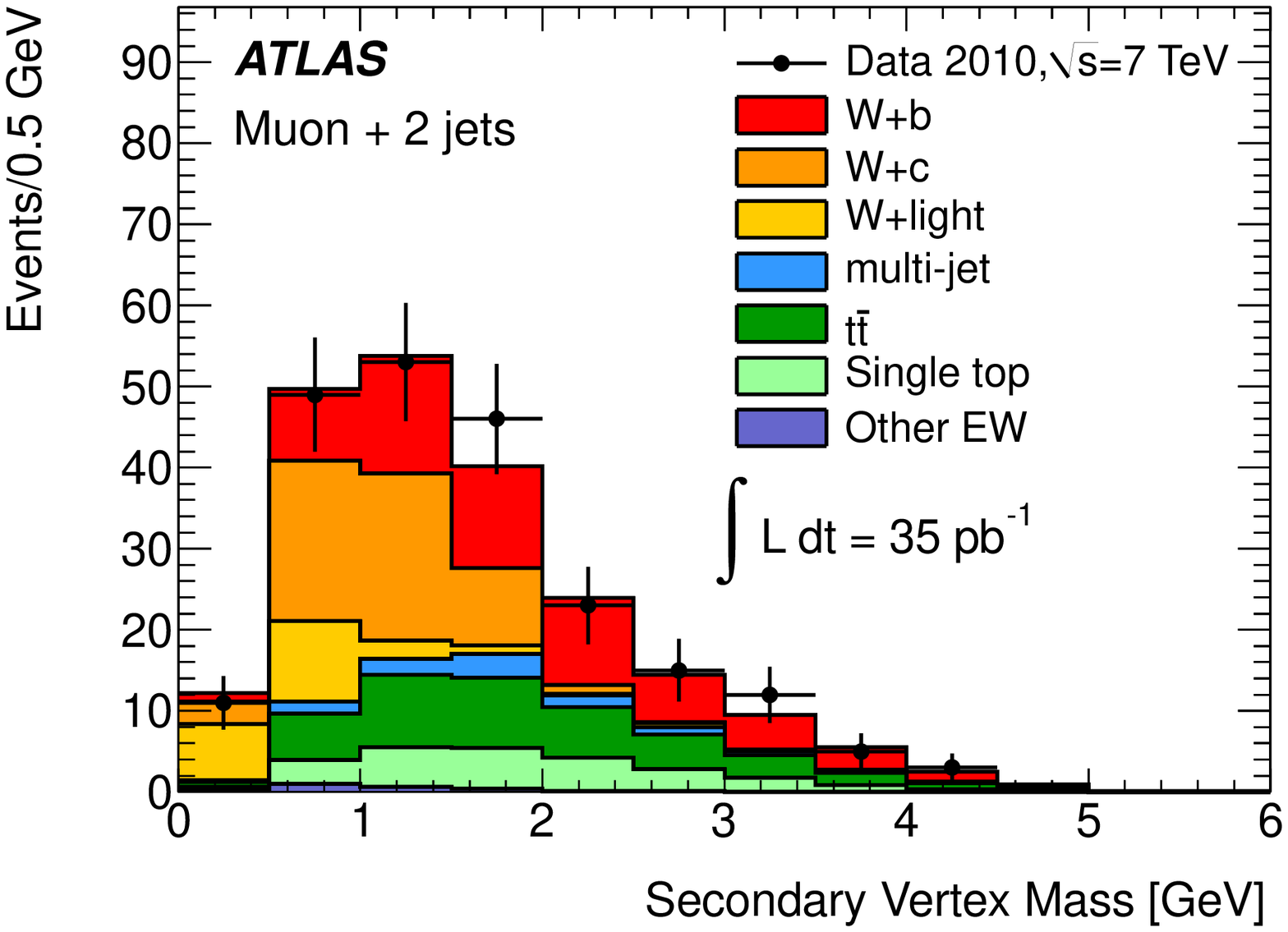} 
   \end{tabular}
   \caption{$m_{SV}$ distributions for the $b$-tagged jet in data and MC,
     where the $W$+jets samples are normalized to the results of the
     maximum likelihood fit and non-$W$+jets backgrounds are normalized
     to the estimates as given in the text, in the 2-jet bin in the
     electron channel (top) and the muon channel (bottom). The stack
     order is the same as in the legend.}
   \label{fig2c}
 \end{figure}
\begin{table*}[ht]
\begin{center}
\caption{Fitted event yields for the $W$+jet contributions including
  the statistical uncertainty from the binned likelihood fit, compared
  to the \textsc{Alpgen} MC prediction normalized to the NNLO inclusive
  $W$ cross section~\cite{FEWZ}, per lepton channel and jet multiplicity. The data-driven multi-jet and
  \ttbar\ estimates and the other background estimates normalized to (N)NLO cross sections are also shown. The statistical uncertainty on the MC prediction is
 negligible and is not shown. No attempt to compare the fitted and predicted
 event yields is made, therefore the systematic uncertainties on
 the predicted event yields have not been estimated. Uncertainties on the MC
 shapes that affect the $W+b$-jets fitted event yield are discussed in section 6.
}
 \begin{tabular}{c||c|c||c|c||c|c||c|c}
\hline\hline
 & \multicolumn{2}{c}{$W\to\mu\nu$, 1-jet}  & \multicolumn{2}{c}{$W\to\mu\nu$, 2-jet}  & \multicolumn{2}{c}{$W\to e\nu$, 1-jet}  & \multicolumn{2}{c}{$W\to e\nu$, 2-jet} \\
\hline
 &  Pred. & Fit result  &  Pred. & Fit result  &  Pred. & Fit result &  Pred. & Fit result  \\ \hline
$W$+$b$   &       25    & $28 \pm 13$ &       26    & $62 \pm 18$ &   $18$ & $33 \pm 12$ & $19$ & $38 \pm 14$    \\
$W$+$c$   &       108   & $170 \pm 20$ &       45    & $54 \pm 19$ &  $84$ & $105 \pm 18$ & $36$ & $24 \pm 15$      \\
$W$+light       &       38    & $21.2 \pm 9.9$ &       20    & $21 \pm 10$ &   $30$ & $22 \pm 10$ & $17$ & $14.4 \pm 7.7$    \\
Multi-jets     &       8       & - &       10    & - &  $10$ & - & $5.8$ & -     \\
\ttbar\   &       11      & - &       44    & - &   $8.1$ & - & $33$ & -    \\
Single top       &       17    & - &       23    & - &  $14$ & - & $18$ &   -   \\
Other backgrounds &       3.9     & - &       2.5     & - &   $1.9$ & - & $2.1$ & -   \\ \hline
Total Predicted &  212 & - & 170 & - & 167 & - & 131 & - \\ \hline
Data &  261 & - & 217 & - & 194 & - & 136 & - \\ 
\hline\hline
\end{tabular}
\label{tab2}
\end{center}
\end{table*}

The $m_{SV}$ distributions in data and MC, where the $W$+jets samples
are normalized to the results of the maximum likelihood fit, are shown
in Fig.~\ref{fig2} and \ref{fig2c}.  The fit results are also shown in
Table~\ref{tab2} and are converted into a $W$+$b$-jet fiducial cross
section times the branching fraction for one lepton flavour, including
corrections for all known detector effects:
\begin{equation}
\sigma^j_{W{\rm + }b{\rm -jet}} \times \mathcal{B}(W \to \ell\nu) = \frac{n^j_{{\rm tag}} \cdot f^j_{W{\rm + }b{\rm -jet}}}{\int L dt \cdot U^j},
\label{eq:XS}
\end{equation}
where the index $j=1,2$ indicated the jet multiplicity,
$n^j_{{\rm tag}}$ the number of selected events with exactly
one $b$-tagged jet, $f^j_{W{\rm + }b{\rm -jet}}$ the fitted fraction of
signal events, $\int L dt$ the integrated luminosity, and
$U^j$ the $W$+$b$-jet correction factor which includes the
acceptance and efficiency effects.

The  correction factor is calculated from the simulation as the ratio of
$W$+$b$-jet events which satisfy the offline selection requirements to
the $W$+$b$-jet events which satisfy the fiducial particle-level
selection criteria, summarized in Table~\ref{tab1}.  At the particle
level jets are reconstructed with the anti-$k_t$ algorithm using all
stable particles ($\tau > 10$~ps), $b$-jets are defined by the
presence of a $b$ hadron with $p_T > 5$~GeV associated to the jet
requiring $\Delta R(\mathrm{jet,}\;b\;\mathrm{hadron}) < 0.3$, and
only weakly-decaying $b$ hadrons are considered. Leptons are defined
by including the energy of all radiated photons within $\Delta R =
0.1$ around the lepton.

The small contributions to the measured $W$+$b$-jet yield from $W\to
\tau\nu$ decays (less than 5\%, where the $\tau$ decays to an electron
or muon) are treated as background and corrected for. The final
correction factors are 0.17 and 0.21 in the electron channel and 0.23
and 0.28 in the muon channel for the 1-jet and 2-jet bin,
respectively. The correction factor is dominated by the $b$-tagging
requirement which has an efficiency of about 35\%. The correction
factor in the electron channel is smaller than in the muon channel due
to tighter electron selection in order to reduce the larger multi-jet
background. Relative uncertainties on the correction factors vary
between 12\% and 14\% and are dominated by the uncertainty on the
$b$-tagging efficiency, as discussed below.

\section{Systematic Uncertainties}
Systematic uncertainties on the measured $W$+$b$-jet cross section are
derived from the non-$W$+jets background estimate, the modelling of
the $m_{SV}$ templates and the correction factor of the fitted
$W$+$b$-jet event distributions to derive the cross section. All
correlations between systematic uncertainties are accounted for.

The largest uncertainty is related to the calibration of the
$b$-tagging efficiency, which impacts not only the $W$+$b$-jet
acceptance and efficiency, but also the template shapes and the
normalisation of the single top background. The uncertainty on the
$b$-tagging efficiency is estimated to be between 6\% for high jet
$p_T>60$~GeV to 13\% at the low $p_T$ end of
25~GeV~\cite{btagcalib}. The uncertainty is driven by the $b$-decay
modelling, the MC statistics, the modelling of the muon $p_T$ spectrum
and the uncertainty on the jet energy scale. The impact of the
$b$-tagging efficiency uncertainty on the \ttbar\ background is
strongly reduced since this background is extracted from data.

The systematic uncertainties on the $m_{SV}$ templates are evaluated
from direct comparisons of the $m_{SV}$ shapes of the data and the
simulation in three multi-jet control regions (an example of the
agreement between data and simulation in such control regions can be
seen in Fig.~19 of Ref.~\cite{btagcalib}). Two of these control
regions are used to determine systematic uncertainties on the bottom
and charm template shapes.  Since both charm and bottom jet tags are
caused by displaced tracks from real vertex decays, it is natural to
determine their uncertainties together from control regions that
enhance the heavy-flavour fractions. One of these control regions is
taken from events in which two jets are $b$-tagged, increasing the
probability that both of the selected jets are from heavy flavour
production.  The other region is taken from $b$-tagged jets which are
also required to contain muons, which is very rare for light-flavour
jets. Both of these control regions are determined to have a
light-flavour contamination of less than 10\%. The bottom and charm
$m_{SV}$ templates used in the $W$+jets fit are then transformed
simultaneously by multiplying by the ratio of the data to the
simulation in the control region for each $m_{SV}$\ bin. The shapes of
the simulated heavy-flavour backgrounds (in particular the top
backgrounds) are also transformed simultaneously. In each lepton
channel, out of the two control regions, the transformation resulting
in the larger variation is chosen to assess the systematic
uncertainty.

Additional studies are performed to account for the possibility that
the charm and the bottom templates may not transform in exactly the
same manner. This is tested by transforming the charm and the bottom
templates one at a time instead of together. It is observed that
varying both the charm and the bottom templates together leads to the
maximum systematic bias, with most of the effect coming from the
distortion of the $b$-template shape, and only about a third of the
effect coming from the distortion of the charm template shape. The
reason that the charm shape plays such a small role in the fit results
is that the template shapes below about $m_{SV} = 1.5$~GeV do not
strongly influence the final fitted $b$-normalisation. The
$b$-normalisation is mostly constrained by the high $m_{SV}$ tail
where there is very little background, especially in the one jet
fits. In fact, fitting the $m_{SV}$ distribution only for $m_{SV} >
1.5$~GeV does not considerably reduce the analysis sensitivity or bias
the final results.

The uncertainty on the measured \ttbar\ yield in the $\ge 4$-jet bin
is dominated by the limited data statistics.  The number of \ttbar\
events is alternatively estimated using a tag-counting
method~\cite{btagcalib}.  The use of simulated \ttbar\ samples for the
projection from the $\ge 4$-jet bin gives rise to systematic
uncertainties from the choice of generator, the amount of QCD initial
and final state radiation (ISR/FSR) and uncertainties on the PDF. The
uncertainty due to the choice of generator is evaluated by comparing
the predictions of MC@NLO with those of
\textsc{Powheg}~\cite{Powheg1,Powheg2,Powheg3} interfaced to either
\textsc{Herwig} or \textsc{Pythia}~\cite{Pythia}. The dominant
uncertainty is represented ISR/FSR, and it is evaluated by studies
using the AcerMC generator interfaced to \textsc{Pythia}, and by
varying the parameters controlling ISR and FSR in a range consistent
with experimental data~\cite{CSCbookISRFSR}. The uncertainty in the
PDFs used to generate \ttbar\ events is evaluated using a range of
current PDF sets with the procedure described in
Ref.~\cite{CSCbookISRFSR}. ISR/FSR and PDF uncertainties are evaluated
in the same way for the single top background.

\begin{figure}[h!]
   \centering
     \includegraphics[width=1.0\columnwidth]{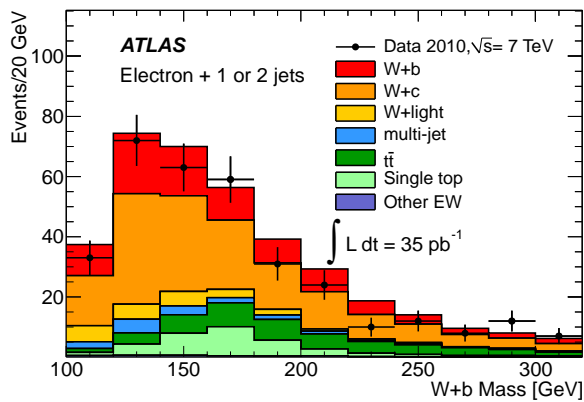} 
     \caption{Invariant mass of the $W$+$b$-jet system in the electron
       channel.  The neutrino $p_z$ is obtained by imposing the $W$
       invariant mass and using the smallest in absolute value of the
       two solutions.  The $W$+jets samples are normalized to the
       results of the maximum likelihood fit and non-$W$+jets
       backgrounds are normalized to the estimates as given in the
       text. }
   \label{fig4}
\end{figure}

Both the \ttbar\ and single top background are irreducible in the
sense that both backgrounds contain a $W$ boson, at least one $b$-jet,
and additional jets. While the \ttbar\ background is extracted from
the data, this is not possible for single top due to the limited
statistics. Therefore, more details are given here on the single top
background. The selection efficiency for single top is considerably
larger than for the $W$+$b$-jet signal, mainly due to the different
$p_T$ spectrum of the $b$-jet. The corresponding single top fiducial
cross sections as defined in Table~\ref{tab1} for one lepton flavour
are 1.4~pb and 1.8~pb in the 1-jet and 2-jet bin, respectively. The
secondary vertex mass shapes for the single top background and
$W$+$b$-jet signal are found to be in good agreement.  The invariant
mass distribution of $W$+$b$-jet in Fig.~\ref{fig4} illustrates good
agreement between data and the fit results.

Uncertainties on the signal modelling are estimated by reweighting the
spectra of both the $b$-jet $p_{T}$ and the opening angle between the
$b\overline{b}$ pairs to match either the \textsc{Herwig} parton
showering or the \textsc{Alpgen} matrix element shapes. The parton
shower model leads to softer $b$-jets and a narrower angle between the
quarks in the $b\overline{b}$ pairs. These modelling uncertainties
affect both the acceptance and efficiency, and the fit templates. It
should be noted, however, that even large changes in the bottom quark
production model have very little effect on the fit template
shapes. The fit template shape dependence on jet kinematics is
weak. The shape also does not depend much on the mode of production
for the heavy-flavour jets except in the rare cases when the two $b$
quarks are produced close to each other such that their fragments are
not resolved in separate jets.  Similarly, even large biases in the
charm quark production kinematics (including varying the rate of $Wc$
production by +/-100\%.) have no significant effect on the fit
template shapes.

The systematic uncertainty on the multi-jet background estimate in the
electron channel is assessed by changing the requirements which define
the control region to model the \met\ template. The uncertainty on the
$m_{SV}$ template shape is estimated in the same way. In addition the
nominal \met\ fit range (0-100~\GeV) is reduced to both 10-100 and
0-60~\GeV. Uncertainties on the EW and top contamination in the
control region are found to be negligible. The uncertainty on the
multi-jet background normalisation in the electron channel is
estimated to be 50\% and is limited by low statistics. The systematic
uncertainty on the muon multi-jet background is dominated by the
validity of the assumptions which go into the matrix method, which is
assessed with closure tests in simulated samples.  The uncertainty on
the $m_{SV}$ template is estimated by an alternative shape
determination using the matrix method bin by bin in $m_{SV}$. The
uncertainty on the multi-jet background normalisation in the muon
channel is estimated to be 30\%. The multi-jet estimate in the muon
channel is further validated by fitting the multi-jet background
explicitly in the $m_{SV}$ template fit by using the muon isolation
variable as a second template. This independent multi-jet estimate
gives consistent results.

The uncertainties on the light jet and $b$-jet energy
scale~\cite{JESv16} as well as the jet energy resolution lead to an
uncertainty on the correction factor for acceptance and efficiency and
to a large uncertainty on the \ttbar\ background normalisation. The
latter is driven by the projection of the measured \ttbar\ yield in
the $\ge 4$-jet bin into the signal region. To a lesser extent
uncertainties on the jet reconstruction efficiency also play a role in
this uncertainty.

Uncertainties related to the lepton trigger and reconstruction
efficiencies are evaluated using tag-and-probe measurements in $Z
\rightarrow ee$\ or $Z \rightarrow \mu\mu$ \cite{incW}. The lepton
momentum scales and resolutions are determined from fits to the
$Z$-mass peak~\cite{incW}.

The missing transverse momentum is recalculated for each systematic
shift applied to the electron, muon, and jet $p_T$.  Additional
uncertainties are applied to soft jets, i.e. those with transverse
momentum below 20~\GeV, and to unassigned calorimeter clusters.  To be
conservative, this uncertainty is considered to be fully correlated
with the uncertainty on the jet energy scale.

A 3.4\% uncertainty on the integrated
luminosity~\cite{bib:Lum,Aad:2011dr} has an impact on the number of
predicted single top, $Z$+jets and diboson events as well as the
conversion from the measured $W$+$b$-jet yield to the cross
section. Uncertainties due to multiple interactions and limited MC
statistics are also considered. Table~\ref{tab3} gives a summary of
all systematic uncertainties.

\begin{table*}[ht]
\begin{center}
\caption{Measured fiducial $W$+$b$-jet cross sections 
  for one lepton flavour with statistical and systematic uncertainty
  and breakdown of relative systematic uncertainties per lepton
  flavour, jet multiplicity, combined across jet bins and also across
  lepton flavour. Uncertainties due to limited MC statistics are
  combined in the template shape uncertainties since this is where the
  low statistics has the biggest impact.}
\begin{tabular}{l|c|c|c||c|c|c||c|c|c}
\hline\hline
 \multicolumn{10}{c}{Fiducial cross section [pb]} \\
\hline
\hline
& \multicolumn{3}{c||}{1 jet} & \multicolumn{3}{c||}{2 jet} & \multicolumn{3}{c}{1+2 jet}\\
            & $\mu$ &  $e$  &$\mu$ \& $e$ &      $\mu$ &  $e$  &$\mu$  \& $e$&      $\mu$ &  $e$  &$\mu$ \& $e$ \\
\hline
     Measured cross section&        $3.5$ &        $5.5$ &        $4.5$ &       $6.2$ &        $5.1$ &        $5.7$ &       $9.7$ &        $10.7$ &        $10.2$ \\
\hline
  Statistical uncertainty&        $1.6$ &        $2.1$ &        $1.3$ &        $1.8$ &        $1.9$ &        $1.3$ &        $2.4$ &        $2.8$ &        $1.9$ \\ 
   Systematic uncertainty&        $1.1$ &        $1.7$ &        $1.3$ &        $1.5$ &        $1.5$ &        $1.4$ &        $2.4$ &        $3.0$ &        $2.6$ \\ 
%Stat. $\oplus$ Syst. &        $1.9$ &        $2.7$ &        $1.8$ &        $2.3$ &        $2.4$ &        $1.9$ &        $3.4$ &        $4.1$ &        $3.2$ \\ 
\hline
 \multicolumn{10}{c}{Breakdown of systematic uncertainty [\%]} \\
\hline
$b$-tag efficiency &                           $15 $ &        $14 $ &        $14 $ &        $10 $ &        $10 $ &        $10 $ &        $11 $ &        $12 $ &        $12 $ \\ 
Template shapes    &                           $16 $ &        $13 $ &        $12 $ &        $10 $ &        $12 $ &        $10 $ &        $11 $ &        $11 $ &        $10 $ \\ 
\ttbar\                &        $9 $ &        $6 $ &        $7 $ &        $12 $ &        $16 $ &        $13 $ &        $11 $ &        $11 $ &        $11 $ \\ 
Single top             &        $10 $ &        $6 $ &        $8 $ &        $4 $ &        $6 $ &        $5 $ &        $7 $ &        $6 $ &        $6 $ \\ 
Signal modeling &        $9 $ &        $8 $ &        $9 $ &        $10 $ &        $10 $ &        $10 $ &        $9 $ &        $9 $ &        $9 $ \\ 
Multi-jets &        $7 $ &        $18 $ &        $11 $ &        $4 $ &        $8 $ &        $4 $ &        $5 $ &        $13 $ &        $7 $ \\ 
Jet uncertainties &        $9 $ &        $6 $ &        $7 $ &        $7 $ &        $10 $ &        $8 $ &        $7 $ &        $7 $ &        $7 $ \\ 
Lepton uncertainties &        $3 $ &        $5 $ &        $3 $ &        $2 $ &        $5 $ &        $3 $ &        $2 $ &        $5 $ &        $3 $ \\ 
\met\ &        $1 $ &        $1 $ &        $1 $ &        $2 $ &        $2 $ &        $1 $ &        $1 $ &        $1 $ &        $1 $ \\ 
   Luminosity &        $5 $ &        $5 $ &        $5 $ &        $4 $ &        $5 $ &        $5 $ &        $5 $ &        $5 $ &        $5 $ \\ 
Multiple interactions &        $5 $ &        $4 $ &        $5 $ &        $3 $ &        $3 $ &        $3 $ &        $3 $ &        $4 $ &        $3 $ \\ 
\hline\hline
\end{tabular}
\label{tab3}
\end{center}
\end{table*}

As a cross check the analysis is repeated using the alternative
JetProb~\cite{JetProb} $b$-tagging algorithm, which gives results
consistent with the default SV0 tagger. The JetProb tagger has a
mistag rate that is more than an order of magnitude higher than the
SV0 tagger and probes a very different mixture of signal and
background.

\section{Results and Conclusions}
The fiducial $W$+$b$-jet cross section in the phase space defined in
Table~\ref{tab1} is measured in the 1- and 2-jet bin in the electron
and muon channel. The results are combined across jet bins and lepton
flavour by summing the corresponding measured cross sections as given
in Equation~\ref{eq:XS}. This linear addition is also performed for
each of the systematic variations considered, in order to properly
take into account the correlations among the different jet bins and
lepton channels due to common systematic uncertainties.  The leading
uncertainties are related to the $b$-tagging calibration and the
$m_{SV}$ template shapes, the top quark background, both \ttbar\ and
single top, the modelling of the signal, the multi-jet background and
the jet energy scale uncertainty. Most of these systematic
uncertainties exhibit a strong correlation with each other between the
jet bins and lepton channels and therefore the relative systematic
uncertainties are only slightly reduced in the combination.

\begin{figure}[h!]
  \centering
  \begin{tabular}{ccc}
    \includegraphics[width=1.0\columnwidth]{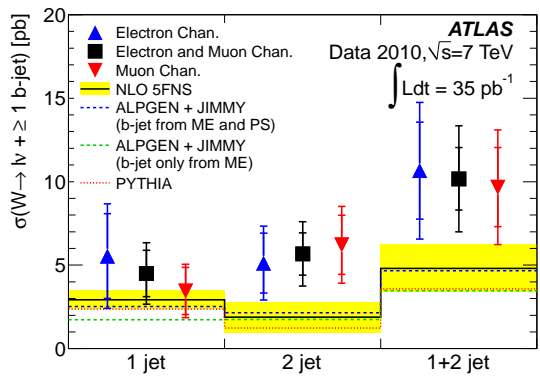}
  \end{tabular}
  \caption{Measured fiducial cross section with the statistical (inner
    error bar) and statistical plus systematic (outer error bar)
    uncertainty in the electron, muon, and combined electron plus muon
    channel.  The cross section is given in the 1, 2, and 1+2 jet
    exclusive bins.  The measurements are compared with
    NLO~\cite{ref:4FNS5FNS} predictions.  The yellow (shaded) band
    represents the total uncertainty on the prediction obtained by
    combining in quadrature the renormalisation and factorization
    scale, PDF set, and non-perturbative correction uncertainties.
    The leading order predictions from \textsc{Alpgen} interfaced with
    \textsc{Herwig} and \textsc{Jimmy} are given for $b$-jets
    generated only by the matrix element and by the matrix element and
    the parton shower. The prediction from \textsc{Pythia} is also
    shown.  }
  \label{fig3}
\end{figure}

The results are presented in Table~\ref{tab3} and Fig.~\ref{fig3} and
are compared with QCD NLO predictions~\cite{ref:4FNS5FNS} performed in
the 5FNS (5 flavour number scheme) described in
Ref.~\cite{ref:4FNS5FNSref,ref:4FNS5FNSref2,ref:4FNS5FNSref3}.  This
calculation requires the combination of two contributions.  The first
contribution has a $b\overline b$ pair in the final state, and the $b$
quarks are considered massive (4FNS).  The second one has a $b$ quark
in the initial state and is treated in a scheme based on $b$ quark
PDFs where the $b$ quark is assumed massless.

The 5FNS prediction is obtained using $\alpha_s^{NLO}(m_Z)=0.118$,
$m_b=4.7~{\rm GeV}$ and $V_{us}=V_{cd}=0.227$. The NLO
CTEQ6.6~\cite{Nadolsky:2008zw} PDF sets are used. The calculation is
obtained with $\mu_R=\mu_F=\mu_0\equiv m_W+2m_b$, where $\mu_R$ and
$\mu_F$ are the renormalisation and factorization scale. The
dependence of the result on the choice of $\mu_0$ is assessed by
varying $\mu$ between $\mu_0/4$ and $4\mu_0$, as in
Ref.~\cite{ref:4FNS5FNS}. These variations account for about a 25\%
uncertainty in the cross section. The PDF uncertainty, estimated to be
at the most 7\%, is obtained by comparing three different PDF sets:
NNPDF2.1~\cite{Ball:2011mu}, CT10~\cite{Lai:2010vv}
MSTW2008~\cite{Martin:2009iq}.

This QCD NLO prediction is only available at the parton level with an
undecayed $W$ boson. The implementation of the NLO 4FNS in
\textsc{Powheg}~\cite{Powheg1,Powheg2,Powheg3} is used to calculate
the $W$ acceptance factor of $0.465\pm0.003$ (stat).  To compare with
data the non-perturbative effects of the hadronization and the
underlying event have to be considered. The impact of these effects
has been evaluated using the \textsc{Pythia} PERUGIA 2011
tune~\cite{perugia} on the \textsc{Powheg} prediction by comparing the
results with hadronization and underlying event model turned on and
off. The non-perturbative correction to the cross section is $0.93\pm
0.07$, dominated by particles from $b$ hadron decays landing outside
the effective anti-$k_t$ jet cone. The systematic uncertainty accounts
for the difference in the modelling of the non-perturbative physics in
\textsc{Pythia} PERUGIA 2011, \textsc{Pythia} MC11, and
\textsc{Herwig}+\textsc{Jimmy} MC11~\cite{mc11}, and the $b$-jet $p_T$
spectrum modelling of \textsc{Powheg}.  Fully corrected predictions
are shown in Table~\ref{tab4} and Fig.~\ref{fig3}. The fiducial
$W$+$b$-jet cross section for one lepton flavour in the combined 1-
and 2-jet bin is measured to be 10.2 $\pm$ 1.9 (stat) $\pm$ 2.6
(syst)~pb. It is found to be larger than the NLO prediction of
4.8$^{+1.2}_{-0.7}$ (scale) $^{+0.3}_{-0.0}$ (PDF) $^{+0.3}_{-0.2}$
($m_b$) $\pm$ 0.3 (non-pert.)~pb, but still consistent within
1.5~$\sigma$.  In addition the measured cross section is compared with
the LO \textsc{Alpgen} prediction of 4.7 $\pm$ 0.1 (stat)~pb, using
the CTEQ6L1~\cite{CTEQ6L1} PDF sets. Here the NNLO correction factor
for the inclusive $W$ cross section of 1.2~\cite{FEWZ} is not applied.

\begin{table}[!htbp] 
\centering 
\caption{Theoretical NLO predictions~\cite{ref:4FNS5FNS} for the $W$+$b$-jet 
  fiducial cross section for one lepton flavour. 
  The systematic
  uncertainties SC, PDF, $m_b$, and NP correspond to the renormalisation and
  factorization scale, PDF set, $b$ quark mass, and non-perturbative correction, respectively, 
  and they are obtained as described in the text.  
}
\begin{tabular}{lllc}
\hline\hline
& Fiducial cross section (NLO) [pb] \\
\hline
1j & 2.9$^{+0.4}_{-0.4}$ SC $^{+0.2}_{-0.0}$ PDF
$^{+0.2}_{-0.1}$ $m_b$  $\pm$ 0.2 NP \\
2j & 1.9$^{+0.8}_{-0.4}$ SC $^{+0.1}_{-0.0}$ PDF
$^{+0.1}_{-0.1}$ $m_b$ $\pm$ 0.1 NP\\
1+2j & 4.8$^{+1.2}_{-0.7}$  SC $^{+0.3}_{-0.0}$ PDF
$^{+0.3}_{-0.2}$ $m_b$ $\pm$ 0.3 NP\\
\hline\hline
\end{tabular}
\label{tab4}
\end{table}

In summary, the cross section for the production of a $W$ boson with
at least one $b$-jet has been measured in $pp$ collisions at
$\sqrt{s}=7$~TeV.  The cross section is measured separately with one
and two associated jets. The results are consistent with the NLO
expectations. The combined result lies above the expectation, but is
consistent at the 1.5~$\sigma$ level.

We are grateful to Laura Reina and Doreen Wackeroth for helpful
correspondence and discussions.  We thank CERN for the very successful
operation of the LHC, as well as the support staff from our
institutions without whom ATLAS could not be operated efficiently.
We acknowledge the support of ANPCyT, Argentina; YerPhI, Armenia; ARC,
Australia; BMWF, Austria; ANAS, Azerbaijan; SSTC, Belarus; CNPq and
FAPESP, Brazil; NSERC, NRC and CFI, Canada; CERN; CONICYT, Chile; CAS,
MOST and NSFC, China; COLCIENCIAS, Colombia; MSMT CR, MPO CR and VSC
CR, Czech Republic; DNRF, DNSRC and Lundbeck Foundation, Denmark;
ARTEMIS, European Union; IN2P3-CNRS, CEA-DSM/IRFU, France; GNAS,
Georgia; BMBF, DFG, HGF, MPG and AvH Foundation, Germany; GSRT,
Greece; ISF, MINERVA, GIF, DIP and Benoziyo Center, Israel; INFN,
Italy; MEXT and JSPS, Japan; CNRST, Morocco; FOM and NWO, Netherlands;
RCN, Norway; MNiSW, Poland; GRICES and FCT, Portugal; MERYS (MECTS),
Romania; MES of Russia and ROSATOM, Russian Federation; JINR; MSTD,
Serbia; MSSR, Slovakia; ARRS and MVZT, Slovenia; DST/NRF, South
Africa; MICINN, Spain; SRC and Wallenberg Foundation, Sweden; SER,
SNSF and Cantons of Bern and Geneva, Switzerland; NSC, Taiwan; TAEK,
Turkey; STFC, the Royal Society and Leverhulme Trust, United Kingdom;
DOE and NSF, United States of America.
The crucial computing support from all WLCG partners is acknowledged
gratefully, in particular from CERN and the ATLAS Tier-1 facilities at
TRIUMF (Canada), NDGF (Denmark, Norway, Sweden), CC-IN2P3 (France),
KIT/GridKA (Germany), INFN-CNAF (Italy), NL-T1 (Netherlands), PIC (Spain),
ASGC (Taiwan), RAL (UK) and BNL (USA) and in the Tier-2 facilities
worldwide.

\bibliographystyle{elsarticle-num}
\bibliography{wb_pub}

\clearpage 
\onecolumn
% ATLAS Collaboration author list for 07-JUL-2011
% Data extracted on 06-Sep-2011 for paperid 107

\begin{flushleft}
{\Large The ATLAS Collaboration}

\bigskip

G.~Aad$^{\rm 48}$,
B.~Abbott$^{\rm 111}$,
J.~Abdallah$^{\rm 11}$,
A.A.~Abdelalim$^{\rm 49}$,
A.~Abdesselam$^{\rm 118}$,
O.~Abdinov$^{\rm 10}$,
B.~Abi$^{\rm 112}$,
M.~Abolins$^{\rm 88}$,
H.~Abramowicz$^{\rm 153}$,
H.~Abreu$^{\rm 115}$
% ...
\bigskip

% ...
$^{10}$ Institute of Physics, Azerbaijan Academy of Sciences, Baku, Azerbaijan\\
$^{11}$ Institut de F\'isica d'Altes Energies and Departament de F\'isica de la Universitat Aut\`onoma  de Barcelona and ICREA, Barcelona, Spain\\
% ...
$^{48}$ Fakult\"{a}t f\"{u}r Mathematik und Physik, Albert-Ludwigs-Universit\"{a}t, Freiburg i.Br., Germany\\
$^{49}$ Section de Physique, Universit\'e de Gen\`eve, Geneva, Switzerland\\
% ...
$^{88}$ Department of Physics and Astronomy, Michigan State University, East Lansing MI, United States of America\\
% ...
$^{111}$ Homer L. Dodge Department of Physics and Astronomy, University of Oklahoma, Norman OK, United States of America\\
$^{112}$ Department of Physics, Oklahoma State University, Stillwater OK, United States of America\\
% ...
$^{115}$ LAL, Univ. Paris-Sud and CNRS/IN2P3, Orsay, France\\
% ...
$^{118}$ Department of Physics, Oxford University, Oxford, United Kingdom\\
% ...
$^{153}$ Raymond and Beverly Sackler School of Physics and Astronomy, Tel Aviv University, Tel Aviv, Israel\\
% ...
\end{flushleft}

\end{document}